\documentclass{article}
\usepackage{natbib}
\usepackage{emulateapj}

\bibpunct{(}{)}{;}{a}{}{,}

\newcommand{\SM}{$M_{\odot}$}

\begin{document}

\submitted{To appear in the Astrophysical Journal 2000}

\title{Numerical Simulations of Globular Cluster Formation}

\author{Naohito Nakasato\altaffilmark{1}, Masao Mori\altaffilmark{2},
and Ken'ichi Nomoto\altaffilmark{1,3}}

\altaffiltext{1}{Department of Astronomy, School of Science,
University of Tokyo, Bunkyo-ku, Tokyo 113-0033;
nakasato@astron.s.u-tokyo.ac.jp, nomoto@astron.s.u-tokyo.ac.jp}

\altaffiltext{2}{Institute of Astronomy, School of Science,
University of Tokyo, Mitaka, Tokyo 181-8588;
mmori@mtk.ios.s.u-tokyo.ac.jp}

\altaffiltext{3}{Research Center for the Early Universe,
School of Science, University of Tokyo, Bunkyo-ku, Tokyo 113-0033}

\begin{abstract}
We examine various physical processes associated with the formation of
globular clusters by using the three-dimensional Smoothed Particle
Hydrodynamics (SPH) code. Our code includes radiative cooling
of gases, star formation, energy feedback from stars including
stellar winds and supernovae, and chemical enrichment by stars.
We assume that, in the collapsing galaxy, isothermal cold clouds
form through thermal condensations and become proto-globular clouds.
We calculate the size of proto-globular clouds
by solving the linearized equations for perturbation.
We compute the evolution of the inner region of the proto-cloud with
our SPH code for various initial radius and initial composition of gases.
When the initial gases contain no heavy elements, 
the evolution of proto-clouds sensitively depends on the initial radius.
For a smaller initial radius, the initial star burst is so intense
that the subsequent star formation occurs in the central regions to form
a dense star cluster as massive as the globular cluster.
When the initial gases contain some heavy elements, 
the metallicity of gases affects the evolution and the final stellar mass.
If the initial radius of the proto-globular clouds was relatively large,
the formation of a star cluster as massive as the globular clusters
requires the initial metallicity as high as [Fe/H] $\geq -2$.
The self-enrichment of heavy elements in the star cluster
does not occur in all cases.
\end{abstract}


\section{Introduction}
The globular clusters belong to the oldest populations in our galaxy.
For the general reviews on globular clusters,
see \citet{Meylan_Heggie_1997} and \citet{Harris_1991}.
Their formation \citep{Elmergreen_1999}
is closely related to the formation process of our galaxy.
The formation of a globular cluster may take place in two stages:
(1) the formation of a proto-globular cloud (PGC) and
(2) the formation of a star cluster from the PGC.
There are three scenarios for the formation of a PGC,
denoted as primary, secondary, and tertiary model \citep{Fall_Rees_1987},
where the PGC forms in different stages, i.e., 
before, during, or after the collapse of the galaxy, respectively.

\begin{itemize}
\item
The primary model was suggested by \citet{Peebles_Dicke_1968}.
They showed that the Jeans mass of the recombination stage of
the universe is comparable to the observed masses of globular clusters
and thus the PGC can form due to the gravitational instability.
Globular clusters may be debris of these objects.
The most serious problem with this primary model is that
there have been few intergalactic globular clusters discovered.
Almost all globular clusters ever discovered exist in galaxies.

\item
In the secondary model, on which we concentrate in the present paper,
the PGC forms due to thermal instabilities \citep{Fall_Rees_1985}.
Our detailed investigation and calculation are presented
in the following sections.

\item
For the tertiary model, one example is that globular clusters
form from large-scale unorganized motion of interstellar gas
as in the Magellanic Clouds,
which are now producing young clusters \citep{Kumai_1993}.
Another example is that very young globular clusters are observed in nearby
galaxy NGC 1705 and 1569 \citep{Ho_Filippenko_1996}.
\end{itemize}

Once the PGC forms, there might be many ways to form a globular cluster (GC).
\citet{Murray_Lin_1993} summarized the scenarios of the formation 
of GCs from PGCs as follows.
The PGC can be divided into two types depending on their masses.
The cloud, whose mass exceeds the Jeans mass, is gravitationally unstable,
thus spontaneously collapsing to form stars.
The cloud, whose mass is smaller than the Jeans mass, is stable until
some instabilities are introduced. A cloud-cloud collision or cloud-disk
collision can trigger such instability.
When such collisions occur, the cloud would become thermally unstable
owing to the efficient cooling. 
This cooling would lead to the formation of a very dense region,
thus inducing a burst of star formation.

In the present paper, we examine quantitatively the above still
qualitative scenario of the globular cluster formation.
We use our Smoothed Particle Hydrodynamics (SPH) code
\citep{Lucy_1977, Gingold_Monaghan_1977} to simulate the
formation of a globular cluster from a PGC.
Our code includes the following physical processes:
radiative cooling, star formation, energy feedback from stars
including stellar winds and supernovae, 
and chemical enrichment from stars.
The SPH method which includes star formation processes like ours 
has been applied to many astrophysical problems.
Such problems include the formation of isolated galaxies
\citep{Katz_1992, Steinmetz_Muller_1994}, 
the evolution of galaxies \citep{Friedli_Benz_1995}, 
and the cosmological simulations \citep{Navarro_White_1993}.
However, our study is the first attempt to apply this method
to the globular cluster formation (preliminary results have been presented in
\cite{Nakasato_1996, Nakasato_1999}).

Among the two triggers of the instability, collapse and collision,
we concentrate on the collapse case in the present paper.
The collision case for a wide range of parameters
(masses of the two clouds, a relative velocity between the clouds etc.)
will be discussed elsewhere.
This paper consists of the following sections.
In section 2 our SPH method is described.
In section 3, we present our linear analysis for the thermal instabilities
to form a PGC.
In section 4, we present the results of the evolution of the PGC with
two different initial compositions of gases.
Sections 5 and 6 are devoted to discussions and concluding remarks,
respectively.

\section{Method}
To simulate the formation of a star cluster from gases,
we use our GRAPE-SPH code using Remote-GRAPE library \citep{Nakasato_1997}.
The GRAPE is a special purpose computer for the self gravity calculations
in general N-body system \citep{Sugimoto_1990}.
Our GRAPE-SPH code includes various physical processes,
e.g., radiative cooling, star formation, and feedback from formed stars. 
There are many different implementations of the SPH method
(a recent comparison of various SPH implementations is presented in 
\citet{Thacker_1998}).
In the present paper, we describe an essential point of our SPH code.
The SPH formulation that we use is the same as the described in
\citet{Navarro_White_1993}.
We use the smoothing length that can vary spatially and evolve with time, 
and integrate equations of motion with a second order Runge-Kutta method
as described in \citet{Navarro_White_1993}.
Details of the implementation of our code are described in \citet{Mori_2000}.
In the following subsections, we describe the physical processes
(cooling, star formation, and feedback) in some detail.

\subsection{Radiative cooling}
The radiative cooling rates depend on the temperature and ionization states
of the gas.
Also, the chemical composition of the gas affects the cooling rate drastically.
We perform SPH simulations for the following three cases (A, B, C) of the gas
with different physical state and chemical composition.

\begin{figure}
\epsfbox[40 28 512 388]{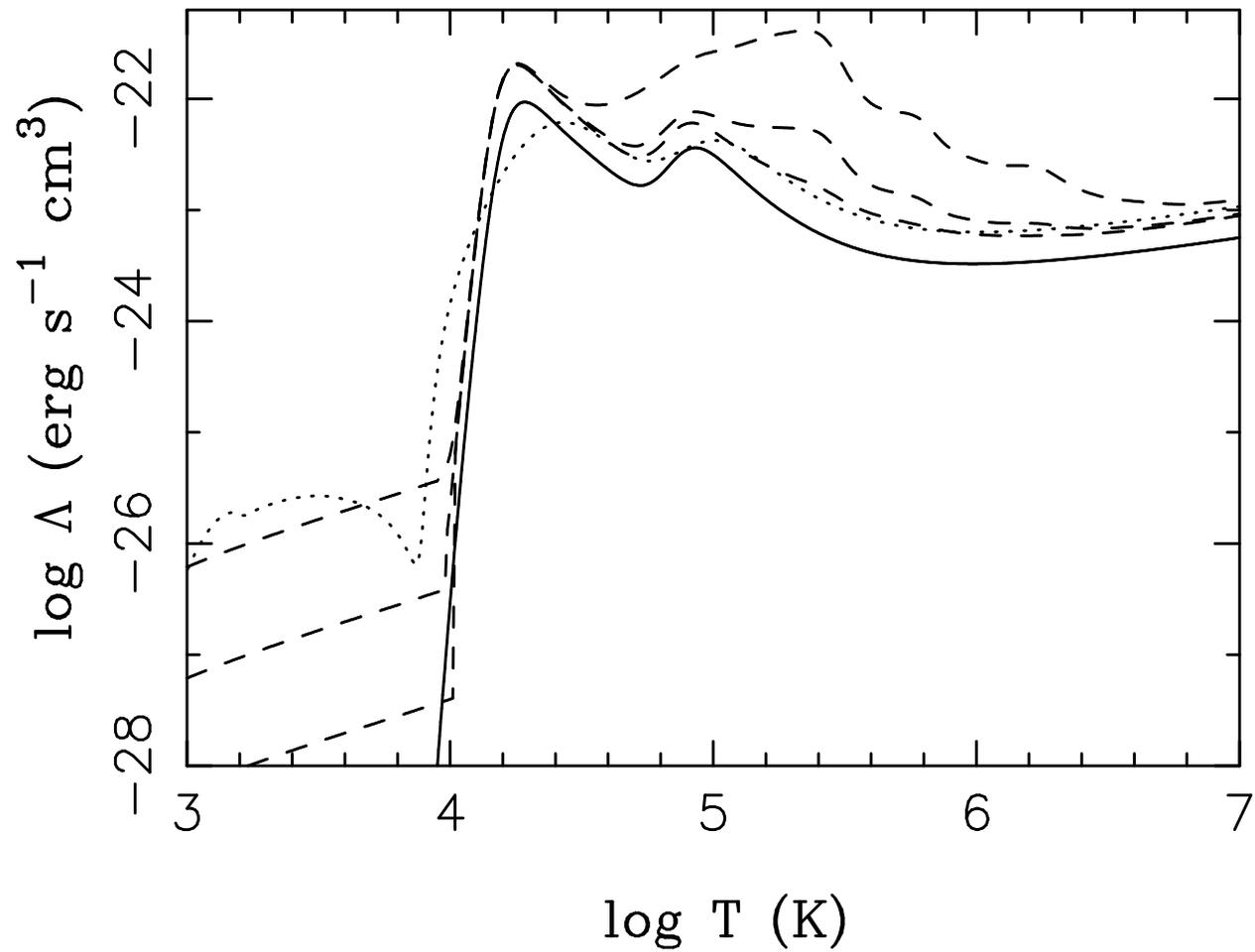}
\caption{
The various cooling rates used in SPH code.
The solid and dotted show the cooling rates for cases A and C, respectively.
The dashed lines show the cooling rates for case B.
From the top to the bottom line, each line corresponds to
[Fe/H] $= -1, -2$ and $-3$, respectively.
\label{cooling}
}
\end{figure}

Case A: We assume that the chemical composition of the gas is primordial
with no heavy elements and the gas is in ionization equilibrium.
In this case, our treatment of radiative cooling is
essentially the same as adopted by \citet{Katz_1995};
they computed the cooling rate using the two-body processes of H and He,
and free-free emissions.
The cooling function ($\Lambda(T)$) is shown in Figure \ref{cooling}
with the solid line.
In this case, the cooling rate decreases very rapidly
as the temperature $T$ decreases below $2 \times 10^4$ K
so that the gas would not radiatively cool below $T \sim 10^4$ K.

Case B: We assume that the gas includes some heavy elements and
is in ionization equilibrium.
In this case, we use the cooling function with different chemical composition
that is computed by MAPPINGS III software by R.S. Sutherland
(MAPPINGS III is the successor of MAPPINGS II that is described in 
\citet{Sutherland_Dopita_1993}).
We compute the cooling function of the ionization equilibrium gas for
[Fe/H] $= -5.0$ - 0.0 with the solar abundance ratio
(see Table 4 of \citet{Sutherland_Dopita_1993}) and 
present the results in Figure \ref{cooling} with the dashed lines
(each line corresponds to [Fe/H] = $-1, -2, -3$ respectively 
from the top to the bottom).
Existence of heavy elements significantly enhances the cooling rates.
At $T < 10^4$ K, the cooling due to the forbidden line emission of
carbon and oxygen is efficient.
For [Fe/H] = $-1$, the cooling rates around $T \sim 10^5$K are
100 times larger than the cooling rates of primordial hydrogen and helium gas. 
These differences would make the evolution of the gas very different.

\begin{figure}
\epsfxsize=0.7\hsize
\epsfbox[44 28 512 392]{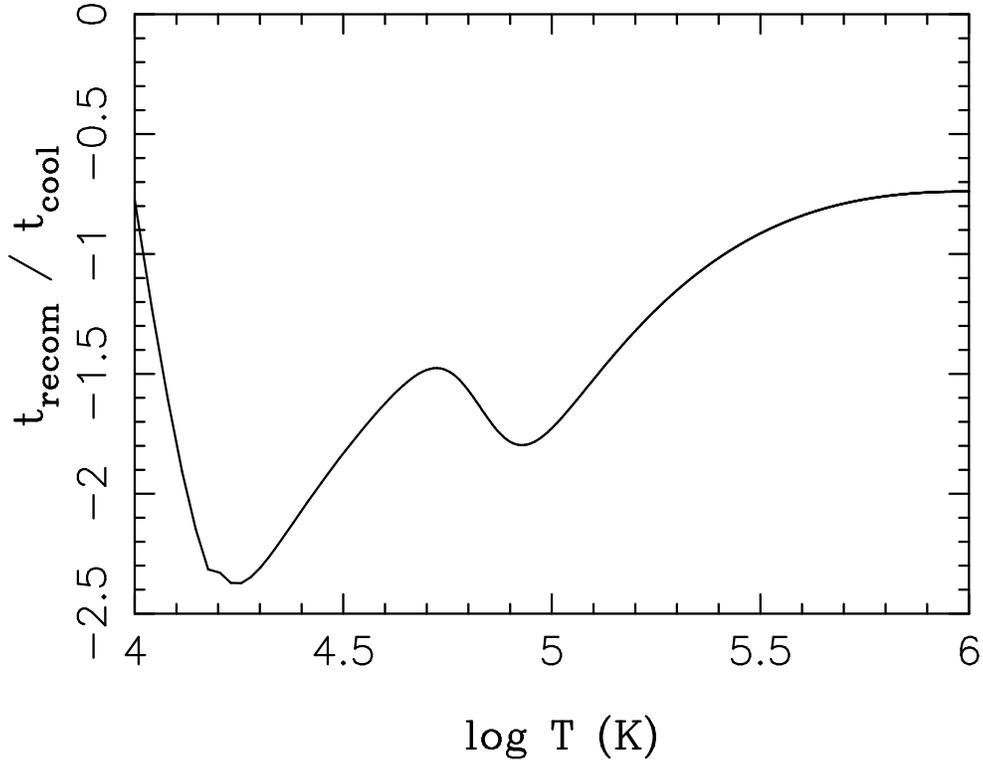}
\caption{
The ratio between the recombination time of hydrogen
and the cooling time ($t_{\rm recom}$/$t_{\rm cool}$)
as a function of the temperature.
\label{recombination}
}
\end{figure}

Case C: We concern about the non-equilibrium cooling.
If the gas cools from high temperatures, 
ionization equilibrium is not realized.
Figure \ref{recombination} shows the ratio between the recombination time 
($t_{\rm recom}$) of hydrogen and the cooling time ($t_{\rm cool}$) of
primordial gas in ionization equilibrium; $t_{\rm recom}$/$t_{\rm cool}$.
The $t_{\rm recom}$ and $t_{\rm cool}$ are defined as
\begin{equation}
t_{\rm recom} = \frac{1.0}{\alpha(T) f_e(T)}
\end{equation}
and
\begin{equation}
t_{\rm cool} = \frac{3 \rm{k_b} \it{T}}{2 \Lambda(T)}, 
\end{equation}
where the $\alpha$ is the recombination coefficient of hydrogen and electron, 
$f_e$ is the fraction of free electrons,
and $\rm{k_b}$ is the Bolzman constant.
For $\alpha$, we use the value of \citet{Spitzer_1978}.
Clearly, the cooling time is much shorter than the recombination time
for $T \sim 2 \times 10^4$ K, so that ionization equilibrium is not realized 
when the gas cools from high temperatures.
In the non-equilibrium case, the existence of electrons and ionized hydrogen
at $T < 10^4$ K makes it possible to form H$_2$ molecules
through the creation of intermediaries H$^-$ and H$_2^+$ as:
\begin{eqnarray}
&\mbox{H} + \mbox{e}  \rightarrow \mbox{H}^- + {\rm photon}&\nonumber\\
&\mbox{H} + \mbox{H}^- \rightarrow \mbox{H}_2 + \mbox{e}&
\end{eqnarray}
and
\begin{eqnarray}
&\mbox{H} + \mbox{H}^+ \rightarrow \mbox{H}^+_2 + {\rm photon}&\nonumber\\
&\mbox{H}_2^+ + \mbox{H} \rightarrow \mbox{H}_2 + \mbox{H}^+&
\end{eqnarray}
\citep{Shapiro_Kang_1987}.
These H$_2$ molecules cause further cooling down to $T \sim 10^2$ K
owing to the rotational-vibrational line excitation.

To include the effect of such molecular cooling, 
we have to solve rate equations that determine the ionization states of 
H and He atoms, and the formation and destruction of H$_2$ molecules.
In our case, the typical time step in solving the rate equations is
shorter than the dynamical time step,
which is determined mainly by the Courant condition.
Integrating all equations with a shorter timestep than
the dynamical time is very costed work.
So we divide the dynamical timestep with dynamical variables (density etc.)
being constant in solving the energy and rate equations, 
which is similar to the method adopted in \citet{Shapiro_Kang_1987}.
The rate-coefficients are also the same as used in \citet{Shapiro_Kang_1987}.
Included species are H$^0$, H$^+$, He$^0$, He$^+$,
He$^++$, H$^-$, H$^+_2$, H$_2$, H$_2^*$ and e, 
where H$_2^*$ is the exited hydrogen molecule.
In our SPH code, we solve the rate equations for 10 species in
each SPH particle with a reasonable computing time.
Solving the rate equations for the gas including heavy elements
(over 200 species) is not feasible with a current resource so that
we only concern the hydrogen and helium plasma in the present paper.

The cooling function for the non-equilibrium case is presented
in Figure \ref{cooling} with the dotted line.
In computing these cooling rates, we follow the isobaric temperature
evolution of the fully ionized gas.
Initially, the temperature and hydrogen number density of the gas
are 10$^7$ K and 0.01 cm$^{-3}$, respectively.
We assume optically thin plasmas so that the gas cools rapidly.
Around $T \sim 2 \times 10^4$ K, H$_2$ molecules begin to form.
At $T < 10^4$ K, the cooling rate due to H$_2$ molecules
is comparable to the cooling rate for [Fe/H] = $-1$ gas.

Finally, we summarize the treatment of the cooling rates in our SPH code.
We can perform SPH simulations for these three cases A, B and C.
In the present paper, we will study the evolution of a PGC with
two different initial chemical compositions, 
e.g., a metal-free gas cloud and a metal-rich gas cloud.
In the former case, we solve the energy and rate equations simultaneously 
for each SPH particle (case B) if the gas temperature is lower than
$3 \times 10^4$ K and the gas particle is not in the heating phase
(see section 2.3).
If these conditions are met, we use the pre-computed cooling table for
the ionization equilibrium case (case A).
In the latter case, we use the pre-computed cooling table to solve the
energy equations (case C).

\subsection{Star formation}
Our treatment of star formation is the same as 
adopted by \citet{Katz_1992}.
Hereafter, ``STAR'' means ``star particle'', 
which is not an individual star but an association of many stars.
A STAR forms in the region where the flow is converging, 
cooling, and Jeans unstable. These conditions are expressed as
\begin{equation}
(\nabla \cdot \mbox{\boldmath{$v$}})_i < 0,
\label{sf_1}
\end{equation}
\begin{equation}
t_{\rm cool} < t_{\rm d},
\label{sf_2}
\end{equation}
\begin{equation}
t_{\rm d} < t_{\rm sound}.
\label{sf_3}
\end{equation}
Here $t_{\rm cool}$, $t_{\rm d}$ , and $t_{\rm sound}$
are the cooling time, dynamical time, and sound crossing time, respectively,
and expressed as
\begin{equation}
t_{\rm cool} = \frac{\mu^2 u}{\rho \Lambda},
\end{equation}
\begin{equation}
t_{\rm d} = \frac{1}{\sqrt{4 \pi G \rho}},
\end{equation}
\begin{equation}
t_{\rm sound} = \frac{h_i}{c_s}, 
\end{equation}
where $\mu$ is the mean molecular weight, $u$ is the specific thermal energy, 
and $c_s$ is the local sound speed.
A STAR forms in the region where these all three conditions are satisfied.
The star formation rate is given as
\begin{equation}
\frac{D \rho_*}{D t} = - \frac{\rho}{t_{\rm starform}}
= - C \sqrt{4 \pi G} \rho^{\frac{3}{2}},
\label{starform}
\end{equation}
where $\rho_*$ is the density of a star, and
$t_{\rm starform} = t_{\rm d} / C$
with a star formation parameter $C$.
In the present paper, we adopt $C = 1.0$,
i.e., a STAR forms in a dynamical time.
The last term of equation \ref{starform} is similar to the Schmidt's law
\citep{Schmidt_1959}.
Integrating equation (\ref{starform}) over one time step $\delta t$
and making calculations with the equations for SPH,
we obtain the mass of a newly formed star in $\delta t$ as
\begin{equation}
m_{\rm star} = \left[ 1 - \frac{1}{1 + 0.5 \frac{\delta t}{t_{\rm starform}}}
\right]
\pi h_i^3 \rho_i.
\label{star_mass}
\end{equation}
The newly formed STAR is then treated as a collision-less particle.
We introduce the minimum allowed mass ($\sim$ 10 \SM) for
the newly formed STAR to prevent the unphysical effects
in the equation of motion and the treatment of the feedback processes.

We note the star formation recipes in our SPH code.
In the regions with increasing densities,
first two conditions for the star formation
(Eq. \ref{sf_1} and \ref{sf_2}) are almost satisfied.
Thus whether STAR forms in some regions is determined mostly
by the Jeans criterion (Eq. \ref{sf_3}).
The critical density for the star formation in our treatment is obtained as
\begin{equation}
\rho_{\rm Jeans} > \left( \frac{\gamma b}{4 \pi G \mu m} \right)^3
\frac{1}{m_i^2} T^3 \left( \frac{1}{\beta^6} \right).
\label{sf_4}
\end{equation}
Here, it is assumed that the neighbor radius is expressed as
\begin{equation}
h_i = \beta \left( \frac{m_i}{\rho_i} \right)^{1/3},
\end{equation}
where $\beta$ is determined experimentally, because $h_i$ is determined
in order to make the number of neighbor particles almost constant
in some range (30 - 80 in our codes).
Since the typical value of $\beta$ is 1.0 - 1.1, $\beta^6$ ranges 1.0 - 2.0.
Then if $m_i$ is constant, the critical density is determined
almost only by the temperature.
The typical calculations in the present paper use 5000 gas particles for
a 10$^6$ \SM~ gas sphere.
Thus the initial mass per particle ($m_i$ in Eq. \ref{sf_4})
is $\sim$ 200 \SM.
For $T = 10^4$ K and 10$^2$ K, the critical density for
star formation is $\rho_{\rm Jeans} \sim 10^{-17}$ and 10$^{-23}$ g cm$^{-3}$,
respectively.
With the star formation recipes used in our SPH code, 
the STAR forms only in the very high density region if $T \sim 10^4$ K.
There are, however, the maximum density ($\rho_{\rm max}$)
that numerically reached in the SPH method;
it is estimated
\begin{equation}
\rho_{\rm max} \sim \frac{N_{\rm n} m_i}{\epsilon^3}, 
\end{equation}
where $N_{\rm n}$ is the number of neighbor particles and $\epsilon$ is the
gravitational softening length.
In the present calculations,
$\rho_{\rm max} \sim 1.8 \times 10^{-20}$ g cm$^{-3}$,
These arguments ensure that with the star formation recipes used
in our SPH code and the initial conditions of the present calculations,
the STAR forms in the region where the temperature is as low as 10$^2$ K.

\subsection{Feedback}
The formed stars eject gases and heavy elements
in stellar winds and Type II supernova explosions
and heats up, accelerate, and enrich
a circumstellar and an interstellar medium.
High energy explosions like supernova produce high temperature and 
low density regions in interstellar medium.
In the SPH method, the numerical accuracy for high density regions
is much better than mesh based methods but the accuracy for
low density regions are poorer.
A typical resolution of usual SPH simulations, 
e.g., \citet{Navarro_White_1993},
including a star formation process (100 - 1000 pc) is
larger than a typical size of supernova remnants ($< 100$ pc).
Thus, it is difficult to properly include the energy, momentum,
and mass release from stars in the SPH method
because of the nature of the SPH method and 
the poor resolutions in current computing resources.
We must use some approximations to mimic real feedback processes
in the SPH method.

One of the method has been proposed by \citet{Navarro_White_1993}.
In their method, the energy produced by a supernova explosion is
distributed to neighbor gas particles of each STAR
mostly as a thermal energy and the rest is distributed
as a velocity perturbation to the gas particles;
the fraction of the energy in a kinetic form is a free parameter
(we note that \cite{Leitherer_1992} presented
the population synthesis model of stellar feedback processes). 
In the present paper, we distribute the energy to neighbor particles 
in a pure thermal form as a zero-th order approximation,
because the size and time scale of our model are much smaller than
those of a galaxy formation model of \citet{Navarro_White_1993}.

\subsubsection{Energy ejection}
The energy ejection rate per STAR is given as
\begin{equation}
E_{\rm eject} = e_{\rm SW} R_{\rm SW} + e_{\rm SNII} R_{\rm SNII}, 
\end{equation}
where $e_{\rm SW}$ is the total energy ejected by stellar winds during
the stellar lifetime and $e_{\rm SNII}$ are the energy ejected
by one Type II supernova explosion.
The $R_{\rm SW}$ is the number of stars per unit time expelling their envelopes
at the current epoch and $R_{\rm SNII}$ is the rate of Type II supernovae.
We define the $R_{\rm SW}$ and $R_{\rm SNII}$ as follows
\begin{equation}
R_{\rm SW} = \frac{\displaystyle{\int^{M_{\rm up}}_{M_{\rm ms}} \Phi(m) dm}}
{\displaystyle{\tau(M_{\rm ms})}}
\end{equation}
\begin{equation}
R_{\rm SNII} = \frac{\displaystyle{\int^{M_{\rm ma}}_{M_{\rm ms}} \Phi(m) dm}}
 {\displaystyle{\tau(M_{\rm ms}) - \tau(M_{\rm ma})}}, 
\end{equation}
where $\Phi(m)$ is the initial mass function (IMF), namely
$\Phi(m) dm$ gives the number of stars in the mass range of ($m$, $m+dm$)
and the $\tau(m)$ is the stellar lifetime \citep{David_1990}.
In the present study, we assume the power law type IMF as
\begin{equation}
\Phi(m) = A m^{-2.35}, 
\end{equation}
where the $A$ is the constant.
For the upper and lower limit masses in Eq. (19),
$M_{\rm up} =$ 120 \SM~ and $M_{\rm lo} =$ 0.05 \SM~ are assumed.
In Eq. (17) and (18), $M_{\rm ma}$ ($=$ 50.0 \SM) and
$M_{\rm ms}$ ($=$ 8.0 \SM) are the upper and lower limit masses
of the stars that explode as Type II supernovae.

For the supernova energy, we assume that $e_{\rm SNII} = 10^{51}$ erg.
For the stellar wind, $e_{\rm SW}$ is estimated to be 0.2$\times 10^{51}$ erg 
for solar metallicity stars from the observational data of OB associations
\citep{Abbot_1982}.
The Chemical abundance of a massive star significantly affects $e_{\rm SW}$ 
\citep{Leitherer_1992}, so that we use metallicity dependent
e$_{\rm SW}$ as $e_{\rm SW} = 0.2 e_{\rm SNII} (Z/Z_{\odot})^{0.8}$, 
where $Z$ is the mass fraction of metal in the STAR.

\subsubsection{Mass ejection}
In our code, the mass ejection due to stellar winds is
combined with the mass ejection due to Type II supernova.
Thus, the mass ejection rate per STAR is written as
\begin{equation}
M_{\rm eject} = m_{\rm SNII} R_{\rm SNII}, 
\end{equation}
where $m_{\rm SNII}$ is the average mass ejected by stellar winds
and Type II supernovae defined as
\begin{equation}
m_{\rm SNII} = \frac{\displaystyle{\int^{M_{\rm ma}}_{M_{\rm ms}} m\Phi(m) dm}}
 {\displaystyle{\int^{M_{\rm ma}}_{M_{\rm ms}} \Phi(m) dm}} - m_{\rm NS}, 
\end{equation}
Here $m_{\rm NS}$ is the mass that is locked up in the neutron star and
assumed to be 1.4 \SM.
The fraction of heavy metal in $M_{\rm eject}$ is computed by the 
nucleosynthesis yield of Type II supernovae
\citep{Tsujimoto_1996, Nomoto_1997}.

\subsubsection{Summary}
We assume that the feedback phase continues for
$\tau(M_{\rm ms}) = 4.3 \times 10^7$ yr from the formation of
each STAR and is divided into two phases:
a stellar wind phase and a supernova phase.
The stellar wind phase continues for $\tau(M_{\rm ma}) = 5.4 \times 10^6$ yr,
during which only the energy ejection from STARs is included;
the ejected mass is included in the supernova phase for simplicity.
The supernova phase begins at $t = \tau(M_{\rm ma})$ and
ends at $t = \tau(M_{\rm ms})$.
During the supernova phase, the energy ejection is given by Eq. (16).
The mass ejection is the sum of the contributions by the stellar winds
and Type II supernovae.

The thermal energies, gases, and heavy elements from stellar winds and
supernovae are smoothed over neighbor particles of
the STAR within a neighbor radius of $R_f$ (feedback radius).
We treat $R_f$ as a parameter to meet the observational constraints.
Such neighbor particles are called ``in heating phase''.
When the gas particles are in heating phase, 
we assume that the cooling is suppressed as proposed in \citet{Mori_1999}.
This treatment produces the high temperature region around the STAR.
Thus, the star formation is forbidden in the gas particles in heating phase.

\section{Proto-Globular Cloud formation}
We first examine the radiative condensations, which occur
in a wide range of astrophysical circumstances
from solar prominence to interstellar clouds \citep{Meerson_1996}.
Radiative condensations in optically thin plasma
have been considered by many authors since the pioneering work
by \citet{Parker_1953} and \citet{Field_1965}.
The scale length of gravitational instability in a collapsing proto-galaxy
is much larger than the radii of globular clusters 
\citep{Lin_Murray_1996}. As will be shown in the following sections, 
the scale length of radiative condensations is comparable
to the radii of globular clusters. Thus, in a collapsing proto-galaxy, 
radiative condensations may be the only mechanism to form a PGC
\citep{Fall_Rees_1985, Lin_Murray_1996}.

\subsection{Radiative condensations in a collapsing proto-galaxy}
The characteristic equation for the growth rate of radiative
condensations, $n$, is obtained from the linearized equations
for perturbations as
\begin{equation}
n^3 + n^2 c_s \left( k_{T} + \frac{k^2}{k_{K}} \right) + n c_s^2 k^2 + 
\frac{c_s^3 k^2}{\gamma} \left( k_{T} - k_{\rho} + \frac{k^2}{k_{K}} \right)
= 0,
\label{chara}
\end{equation}
where $c_s$ is the sound speed, $\gamma$ is the ratio of the specific heats, 
$k = 2 \pi / \lambda$ is the wavenumber of the perturbation, 
$k_{\rho}$ and $k_{T}$ are the wavenumber of sound waves
whose frequencies are equal to the growth rate of isothermal
and isochoric perturbation,
respectively, and $k_{K}$ is the inverse of 
the scale length of thermal conduction \citep{Field_1965}.
$k_{\rho}$, $k_{T}$ and $k_{K}$ are expressed as
\begin{equation}
k_{\rho} = \frac{\mu (\gamma - 1) \rho_0}{c \rm{k_b} T_0}
\frac{\Lambda(T_0)}{\mu^2}, 
\end{equation}
\begin{equation}
k_{T} = \frac{\mu (\gamma - 1) }{c \rm{k_b}}
\frac{\rho_0}{\mu^2} \frac{d \Lambda}{d T},
\end{equation}
\begin{equation}
k_{K} = \frac{c \rm{k_b} \rho_0}{\mu (\gamma - 1) \kappa},
\end{equation}
where $T_0$ and $\rho_0$ are the equilibrium temperature and density,
respectively, k$_b$ is the Bolzman constant,
and $\Lambda(T)$ is the cooling function (Figure \ref{cooling}).
We assume $\gamma = 5/3$ and $\kappa = 5.6 \times10^{-7} \quad T^{2.5}$ 
erg s$^{-1}$ K$^{-1}$ cm$^{-1}$.
Solving Eq. (\ref{chara}) as a cubic equation of $n$ for different $k$,
we obtain the dispersion relation between $n$ and $k$.
In applying to our galaxy, we adopt $T_0 \sim 1.0 \times 10^6$ K and
$\rho_0 \sim 1.7 \times 10^{-24}$ g cm$^{-3}$ \citep{Fall_Rees_1985}.
With these values, the dispersion relation has a peak at some $k$ .
In the present paper, we assume that the scale for the maximum growth rate
is typical scale of a PGC.
To obtain the typical scale of a PGC for different $T_0$ and $\rho_0$, 
equation (\ref{chara}) is viewed as a quadratic in $k^2$
(see Section II (d) in \citet{Field_1965}).

The results for different $T_0$ and $\rho_0$ are presented in Figure \ref{m_n}.
In applying to our galaxy, we obtain the scale length of  $\sim 600$ pc.
This scale length is consistent with the following simple estimate.
For the adopted cooling function, the wavelengths for $k_{\rho}$ and $k_{K}$
with $T_0 \sim 1.0 \times 10^6$ K and
$\rho_0 \sim 1.7 \times 10^{-24}$ g cm$^{-3}$
are respectively obtained as
\begin{equation}
\lambda_{\rho} > 10^3 \quad \mbox{pc}
\quad \mbox{and} \quad \lambda_{K} < 1 \quad \mbox{pc}.
\end{equation}
This implies that the perturbation with a scale greater than $10^3$ pc
is dumped owing to the limit of the sound speed, and the perturbation
with a scale smaller than 1 pc is also dumped by thermal conduction.
Thus the typical scale of a PGC in our galaxy is estimated to
be several hundreds pc and the mass of a PGC ranges
from 10$^7$ to 10$^8$ \SM.

\subsection{Density profile of a PGC}
The estimated scale of a PGC is larger than the present radius of
globular clusters (10 - 100 pc).
This implies that during radiative condensations, 
a PGC shrinks before star formation begins or star formation in a PGC
occurs in the central region of the cloud.
If we consider the metal-free PGC where the ionization equilibrium is achieved,
the cooling time, $t_c$, of the PGC is much shorter than the dynamical time,
$t_d$ ($t_c/t_d \sim $ 0.25 for $T_0 \sim 1.0 \times 10^6$ K
and $\rho_0 \sim 1.7 \times 10^{-24}$ g cm$^{-3}$).
After the short period of radiative condensations ($\sim t_c$),
the PGC becomes a warm dense cloud ($T \sim 10^4$ K) surrounded by
a hot thin gas with $T_0$ \citep{Fall_Rees_1985}. 

The qualitative discussion on the evolution of such a metal-free PGC
has been presented by \citet{Lin_Murray_1996}.
They argued that in the first stage of collapsing galaxy,
the assumption of ionization equilibrium is not valid
because of little radiative emission. 
If the ionization equilibrium is not achieved, the temperature of 
high density regions can be as low as $\sim 10^2$ K due to hydrogen molecular
cooling. In this case, the high density cold cloud with $T = 10^2$ K is
surrounded by a warm gas.
The critical mass of the isothermal sphere confined by
an external pressure depends on the cloud temperature
\citep{Ebert_1955, McCrea_1957}.
If the temperature of the cloud is $\sim 10^2$ K, 
the critical mass is $\sim 10^2$ \SM.
According to \citet{Lin_Murray_1996}, such a small cloud as
$M \sim 10^2$ \SM~ collapses
to make the first stars in the proto-galaxy.
These first generation stars are the source of radiative
emission to maintain the ionization equilibrium.
If such radiation continues long enough, a cold PGC will evolve
almost isothermally with $T \sim 10^4$ K and form the isothermal
profile of $\rho \propto r^{-2}$.
The core size of the PGC will decrease from the initial size of
several hundreds pc to a size comparable to the observed globular cluster.
According to these arguments, we can assume that the density structure of a PGC
has a form of $\rho \propto r^{-2}$.

\begin{figure}
\epsfbox[40 28 524 388]{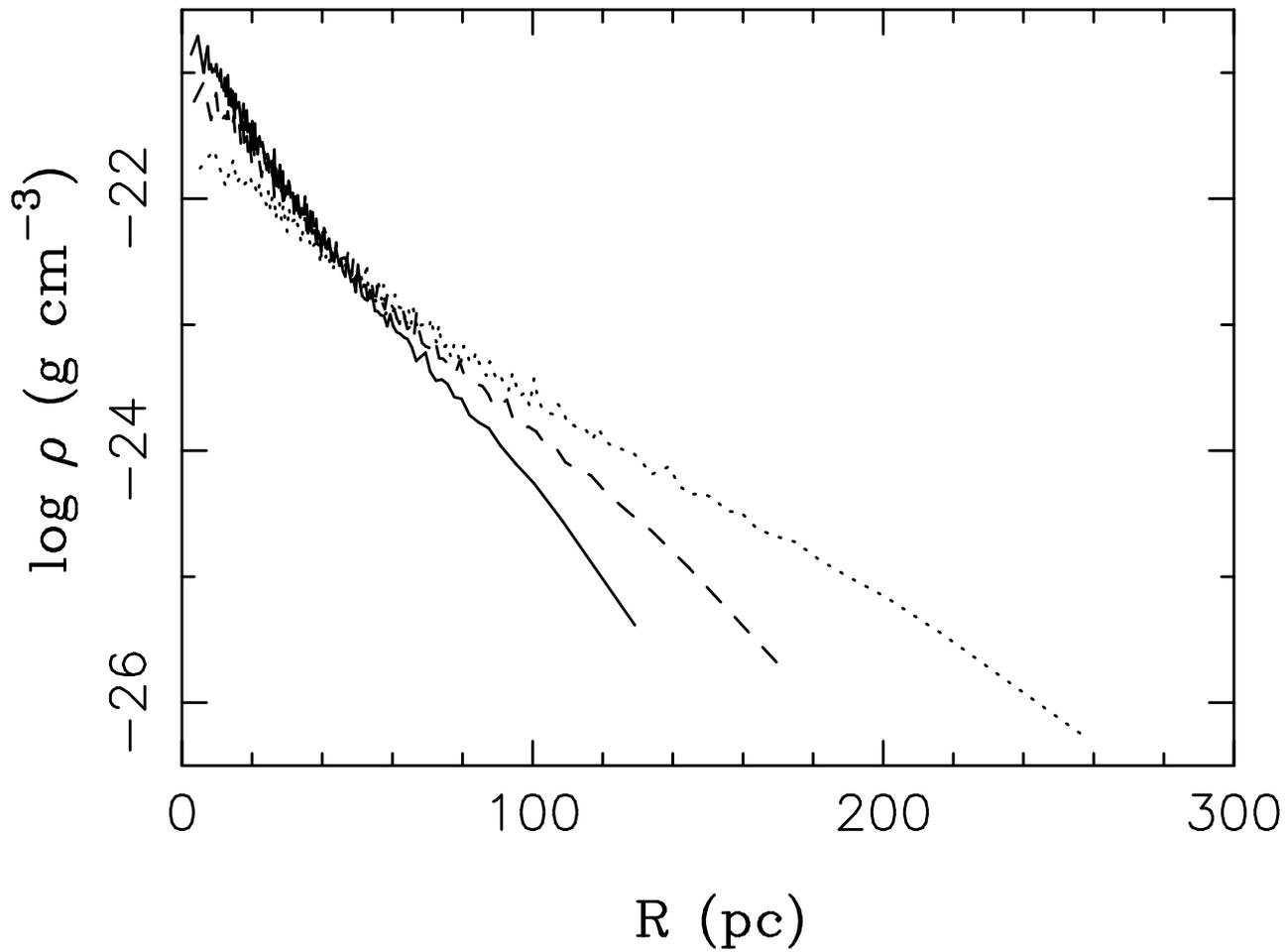}
\caption{
The initial gas density (g cm$^{-3}$) profile for the King profile case.
The solid, dashed, and dotted lines show the profiles
for $R_i =$ 150, 200, and 300 pc, respectively.
\label{initial}
}
\end{figure}

For the initial conditions of the PGC,
we use the King profile with fixed mass of 
$M \sim 10^6$ \SM~ which nearly equals to the Jeans mass of the
current conditions ($\rho_0$ and $T_0$).
We assume that this sphere represents the inner regions of the PGC.
We examine three different cases for a initial radius of
$R_i =$ 150, 200, and 300 pc.
The initial gas density profiles of the three cases are presented
in Figure \ref{initial}.
For the smaller $R_i$,  the initial concentration of the inner region of
the PGC is higher.
The temperature of the sphere is assumed to be $T \sim 10^4$ K.
The properties of the inner region of the PGC, which are
the initial conditions of our calculations, are summarized as follows:
\begin{itemize}
\item
The mass of the clouds is $M \sim 10^6$ \SM.

\item
The radius of the clouds is $R_i \sim $ 150 - 300 pc.

\item
The clouds is an isothermal sphere with $T \sim 10^4$ K.

\item
Initially, the velocity of the cloud is zero.
\end{itemize}
In the following calculations, the initial number of gas particles 
is $\sim 5000$ so that the initial mass of the one gas particle is 
$\sim$ 200 \SM.
The dependence of varying the initial number of gas particles
is discussed in Section \ref{parameter}.
In all cases, the gravitational softening length for all particles
is set to be 1 pc and fixed during the calculations.

\section{Results} 
First, we note on the feedback radius ($R_f$).
We use the Str\"omgren radius ($R_{\rm St}$) of a typical OB star as $R_f$.
The typical value is $R_{\rm St}\sim$ 10 - 100 pc,
where the density of the ISM is $\sim$ 1 cm$^{-3}$ \citep{Osterbrock_1974}.
The radius depends on the density of the ISM as $R_{\rm St} \propto n^{-2/3}$.
In the central region of our initial models,
$n$ ranges from 100 cm$^{-3}$ to 1000 cm$^{-3}$.
Thus, we can estimate $R_f$ $\sim 1 - 5$ pc
(using the largest value of $R_{\rm St}$).
In the following two subsections, we choose $R_f =$ 3 pc for all gas particles.
The results for both cases are summarized in Table \ref{table1}.
The results for different $R_f$ are described in Section \ref{parameter}.

\subsection{Evolution of the metal-free PGC}
In this section, we describe the evolution of the metal-free PGC for 
case C.

As presented in the previous sections, the metal-free PGC may
evolve isothermally with $T \sim 10^4$ K.
The isothermal evolution of PGC is terminated
when the ionizing radiations from the first generation stars stop.
Without ionizing radiation, the PGC cools efficiently by H$_2$ molecules.
Our calculations start from that time when the PGC begins to cool.
We assume $M = 10^6$ \SM, and the initial temperature of
$T = 10^4$ K and the King profile sphere.
Since the size of the PGC at this stage is unknown, 
we examine three different cases for an initial radius of
$R_i =$ 150, 200, and 300 pc.

In all three cases, the initial density of the central region is high enough
for efficient cooling to occur, so that the temperature of the central region
decreases rapidly to $\sim 10^2$ K.
Due to the high density and low temperature ($\sim 10^2$ K),
a burst of star formation occurs in the central region in all cases.
After the first star formation, the central region becomes the heating region
and the temperature of the central region increases gradually.

\begin{figure}
\epsfbox[37 305 538 531]{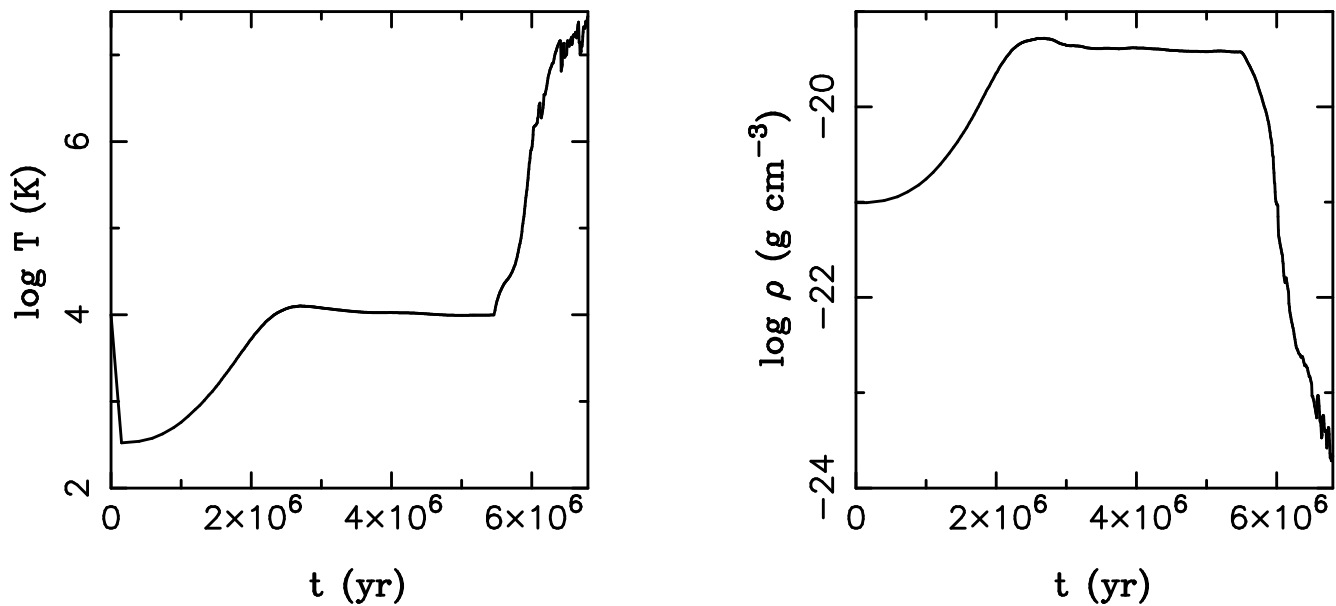}
\caption{
The evolutionary changes in the central temperature (left panel) and
gas density (right panel) are shown for the metal-free collapse
with $R_i =$ 150 pc.
\label{150_free_t_d}
}
\end{figure}

\begin{figure}
\epsfbox[34 305 548 531]{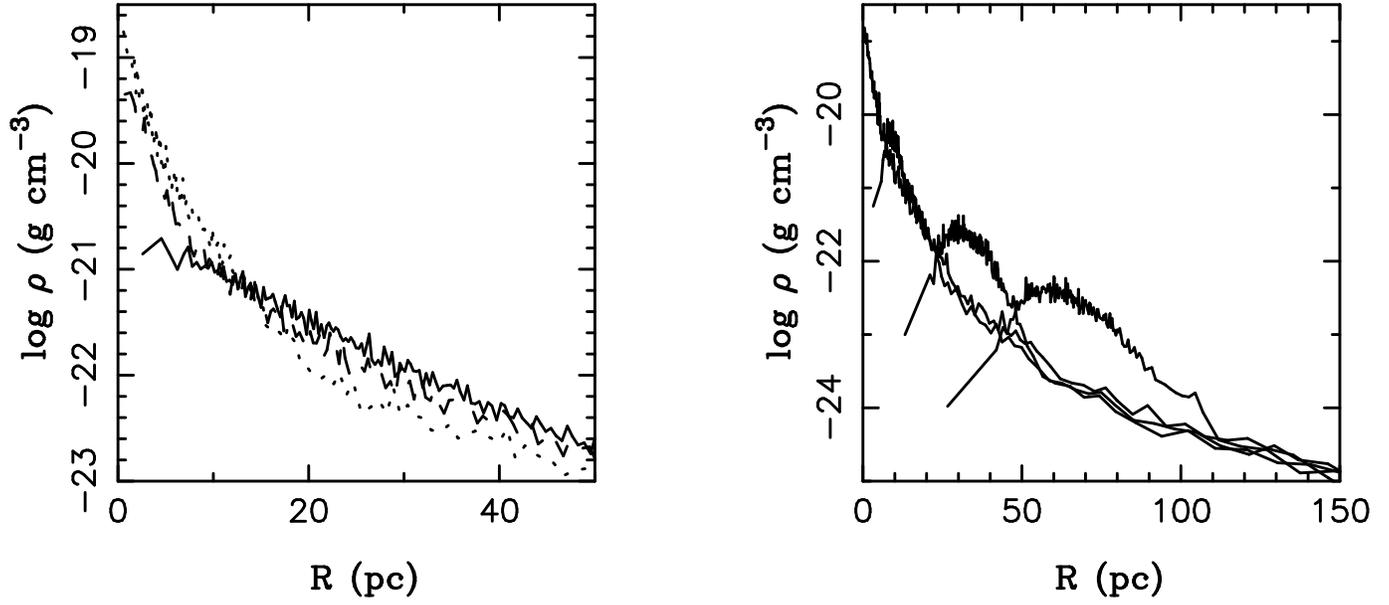}
\caption{The gas density profile as a function of radius (pc).
left: The solid, dashed and dotted lines respectively shows the profile
at $t =$ 0, 2, 4 Myr.
right: The solid lines with the decreasing central density
correspond to $t$ = 5, 6, 7, 8 Myr.
\label{150_free_rho}
}
\end{figure}

At $t \sim$ 6 Myr, Type II supernovae start to occur and the temperature of the
heating region increases rapidly to T $\sim 10^6$ K.
The left panel of Figure \ref{150_free_t_d} shows
the evolutionary changes in the central temperature for $R_i$ = 150 pc.
Such a high temperature region expands and
the central density begins to decrease
(see the right panel of Figure \ref{150_free_t_d}).
The expansion of the central region leads to the formation of
a shell structure as seen in the evolution of the gas density profile
(Figure \ref{150_free_rho}).
The density profile at $t =$ 6 Myr, clearly shows the shell structure.
After that time, the shell expands outwardly
(see the right panel of Figure \ref{150_free_rho}).

\begin{figure}
\epsfbox[37 305 538 541]{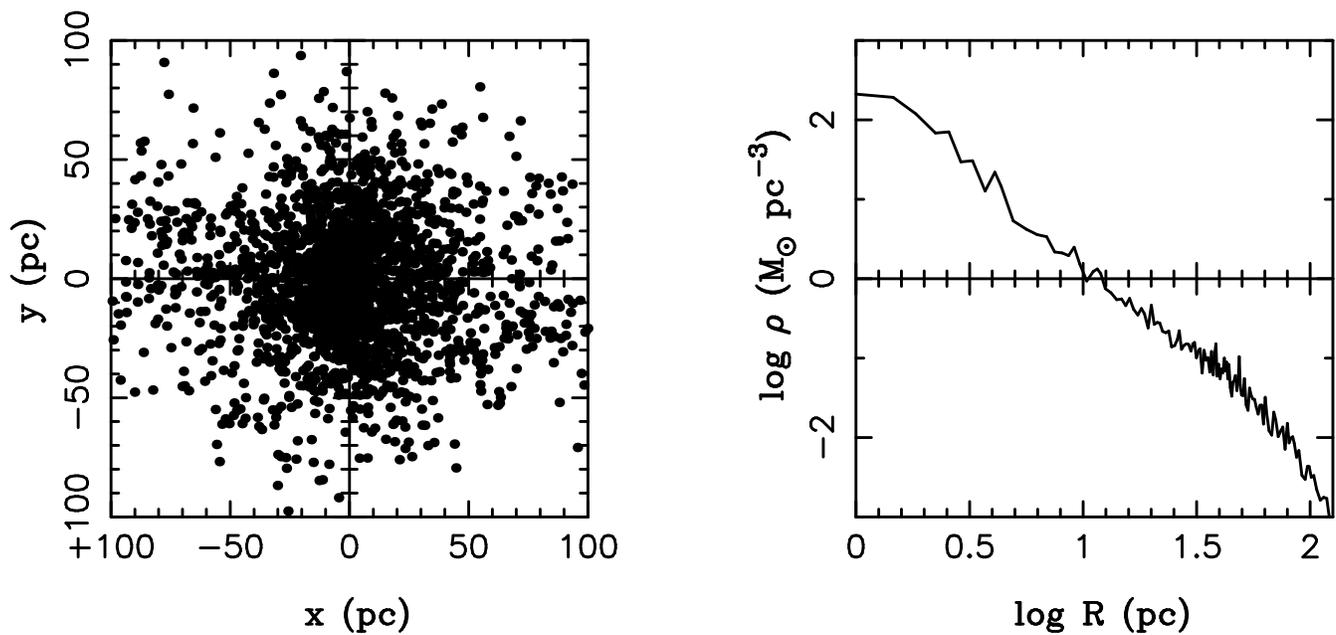}
\caption{
The projected position of particles at $t =$ 10 Myr
for metal-free collapse with $R_i =$ 150 pc (left panel).
The stellar density profile at $t =$ 10 Myr for the same model.
\label{150_free_p}
}
\end{figure}

The star formation after the shell formation occurs not in the central region
but in the shell.
Figure \ref{150_free_birth} show the change in the radius of
the star forming region as a function of time for $R_i$ = 150 pc.
We can see that the star forming region moves outward.
At $t =$ 10 Myr, where we stop the computation for $R_i =$ 150 pc,
the stellar mass reaches $\sim 1.3 \times 10^5$ \SM.
This means that the star formation efficiency is $\sim$ 13 \%.
The bound stellar mass at $t =$ 10 Myr is $\sim 10^5$ \SM~ and
the mass is comparable to the typical mass of globular clusters
($\sim 10^5$ \SM).
At $t =$ 10 Myr, the gas is almost removed from the central region.
Figure \ref{150_free_p} shows the projected particle positions
of STAR particles (left panel) and the stellar density profile (right panel)
at $t =$ 10 Myr.
The STAR particle shows elongated shape (bar like shape).
Such a shape is caused by the radial orbit instability of the STARs formed
at the large radius \citep{Palmer_Papaloizou_1987}.
The central density of STARs is as high as $\sim$ 100 \SM~ pc$^{-3}$
and the central velocity dispersion of the STARs is $\sim 3 $ km s$^{-1}$.
This value of velocity dispersion is smaller than the observed value
\citep{Dubath_1997}.
We need to follow further evolution of the star clusters
for proper comparison, which is not feasible with present code.
The number of STAR particles at $t =$ 10 Myr is $\sim$ 2500.

For $R_i =$ 200 pc and 300 pc, the overall evolution is similar to the case
for $R_i =$ 150 pc and the stellar masses at $t =$ 10 Myr are
$\sim 8.5 \times 10^4$ \SM~ and $\sim 5.6 \times 10^4$ \SM,
respectively.
The bound stellar mass are $\sim 5 \times 10^4$ \SM~ for both case.
These results are summarized in Table \ref{table1}.
The initial concentration correlates with the final stellar mass
and the final concentration of stellar system.
In order for the stellar system as massive as 10$^5$ \SM~ to form, 
the PGC should be as compact as $R_i < 200$ pc
($\rho_{\rm c} > 10^{-22}$ g cm$^{-3}$) for the metal-free condition.

\subsection{Evolution of the metal-rich PGC}
The chemical composition of stars in globular clusters is one of the
most crucial quantities to constrain the model
for the globular cluster formation.
The assumption that PGC initially has some heavy elements is quite reasonable.
Actually, all globular clusters in our galaxy have some metals
of [Fe/H] $= -2.25$ - 0.
The metallicity distribution of the globular clusters shows
bimodal distributions \citep{Harris_1991}.
From this fact, we choose the initial metallicity of the cloud ranging
from [Fe/H] $= -2$ to 0.
For comparison, we will evolve the lower metallicity ([Fe/H] $= -3$) cloud.
The results presented in this section are obtained for case B.

\begin{figure}
\epsfbox[37 305 543 540]{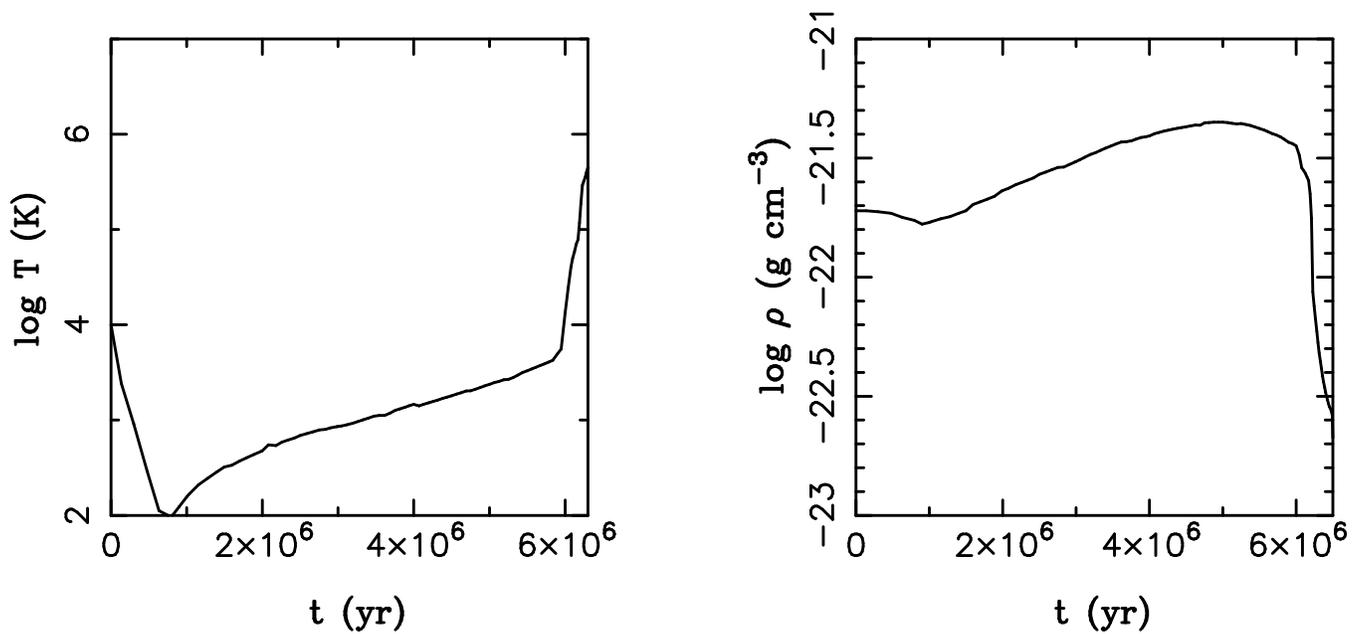}
\caption{
The evolutionary changes in the central temperature (left panel) and
gas density (right panel) are shown for the collapse of [Fe/H] = $-2$ sphere.
\label{300_m20_t_d}
}
\end{figure}

Using the same initial conditions presented in section 4.1, 
we evolve the PGC for different metallicity.
For the initial radius of the cloud, we set $R_i =$ 300 pc.
The first star formation occurs in the central region and the STARs heat
up the surrounding matter to gradually increase
the temperature of the central region.
At $t \sim$ 6 Myr, Type II supernovae begin to occur and
the temperature of the heating region increases rapidly to $T \sim 10^6$ K.
Figure \ref{300_m20_t_d} shows the evolutionary changes
in the central temperature (left panel) and gas density (right panel)
for [Fe/H] $= -2$.
The evolution for [Fe/H] $= -2$ is very similar to
the evolution of the metal-free cloud.
After $t \sim$ 6 Myr, the central high density region makes the shell 
structure as in the metal-free case.
At $t =$ 10 Myr, the bound stellar mass reaches $\sim 1.0 \times 10^5$ \SM~
for [Fe/H] $= -2$.

\begin{figure}
\epsfxsize=0.7\hsize
\epsfbox[44 28 518 388]{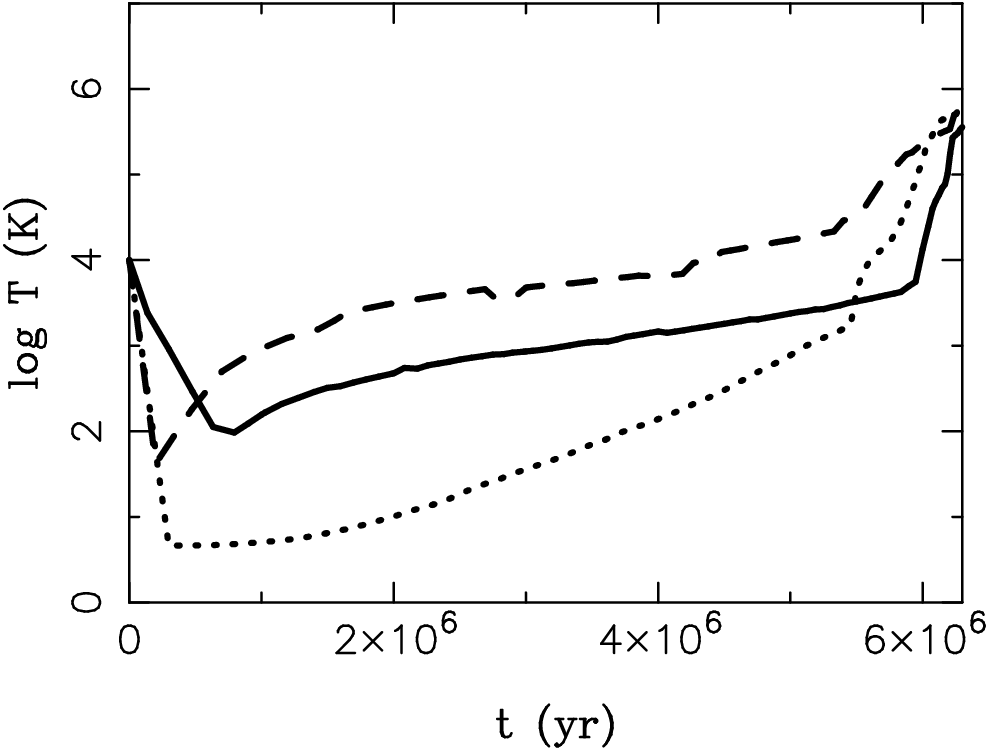}
\caption{
The evolutionary changes in the central temperature for the collapse
of [Fe/H] $= -2$ (solid line), $-1$ (dashed line)
and $0$ (dotted line) sphere.
\label{metal_t}
}
\end{figure}

\begin{figure}
\epsfxsize=0.7\hsize
\epsfbox[40 28 509 402]{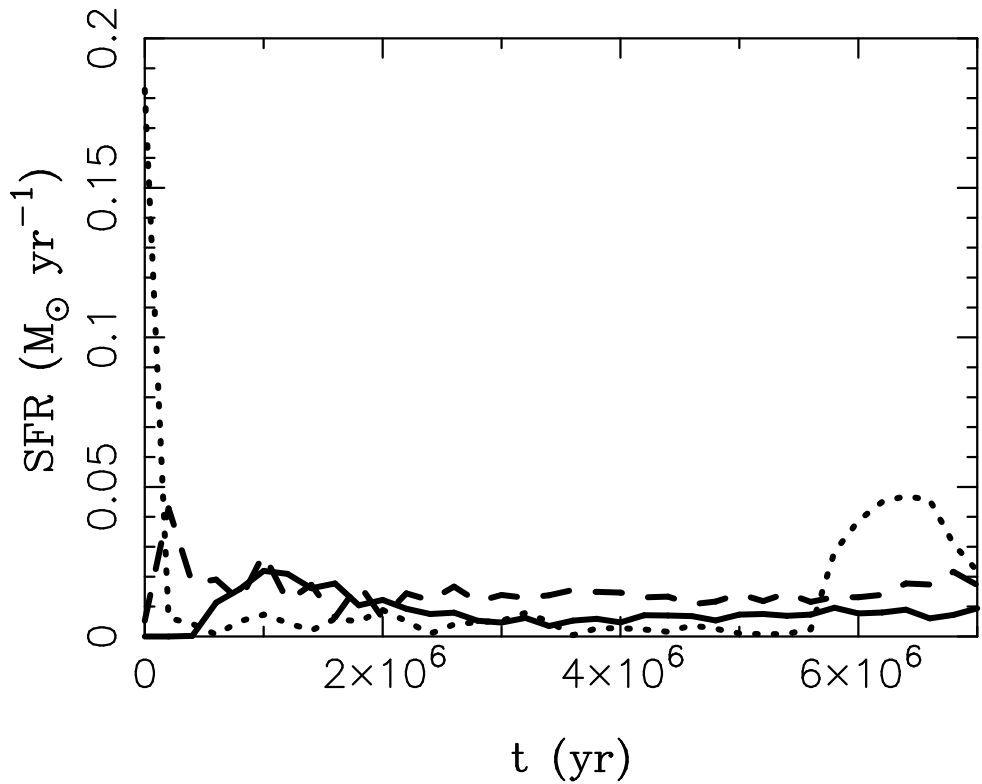}
\caption{
The comparison of SFR as a function of time for 
[Fe/H] $= -2$ (solid line), $-1$ (dashed line) and $0$ (dotted line).
\label{metal_sfr}
}
\end{figure}

For higher metallicity, i.e., [Fe/H] $\geq -1$, 
details of the evolution are somewhat different.
We compare the central temperature evolution for different metallicity
in Figure \ref{metal_t}.
Due to the different cooling rate, the initial decrease in
the temperature is larger for higher metallicity.
This difference leads to different star formation history as shown in
Figure \ref{metal_sfr}.
For [Fe/H] $= 0$, the initial star burst is very intense and then
the SFR decrease sharply becoming lower than the low metallicity case
after $t =$ 1 Myr.
This is because heating due to the stellar winds much larger
for higher metallicity (see Section 2.3.1).
The rise in the SFR after $t =$ 6 Myr for [Fe/H] = 0
is caused by Type II supernovae.

The SFR for [Fe/H] $= -1$ is almost constant during the evolution.
For [Fe/H] $= -2$, the first star formation occurs at $t \sim$ 0.5 Myr
because of the lower cooling rates.
The SFR after $t =$ 2 Myr is almost constant but lower than
for [Fe/H] $= -1$.

The bound stellar mass at $t =$ 10 Myr is $\sim 1.5 \times 10^5$ \SM~
and $1.0 \times 10^5$ \SM~ for [Fe/H] $= -1$ and 0, respectively.
The reason for such a metallicity dependence is summarized below.
The bound stellar mass, the central stellar density, and the
central velocity dispersion are summarized in Table \ref{table1}.
The results depend on the metallicity as follows:

\begin{enumerate}
\item
The initial metallicity  significantly affects the star formation history.
This difference makes the subsequent evolution different.

\item
For the higher metallicity, the final stellar mass at $t =$ 10 Myr
is larger because of the more efficient cooling rate.

\item
The bound stellar mass is not an increasing function of
the initial metallicity.
This is because heating due to the stellar winds is
larger for higher metallicity.

\item
For lower metallicity ([Fe/H] $= -3$), only $\sim 3 \times 10^3$ \SM~
stars form.
This implies that [Fe/H] $\geq -2$ is necessary to form
globular clusters of $\sim 10^5$ \SM~ if $R_i \sim$ 300 pc.
\end{enumerate}

\subsection{Parameter dependence}
\label{parameter}

In this subsection, we describe the dependence of the results
on various numerical parameters, e.g., the initial particle number
($N_i$) and the feedback radius ($R_f$).
We use the metal-rich cloud of [Fe/H] $= -1$ as a reference model,
where $N_i =$ 5000 and $R_f =$ 3 pc.

\begin{figure}
\epsfbox[40 28 524 407]{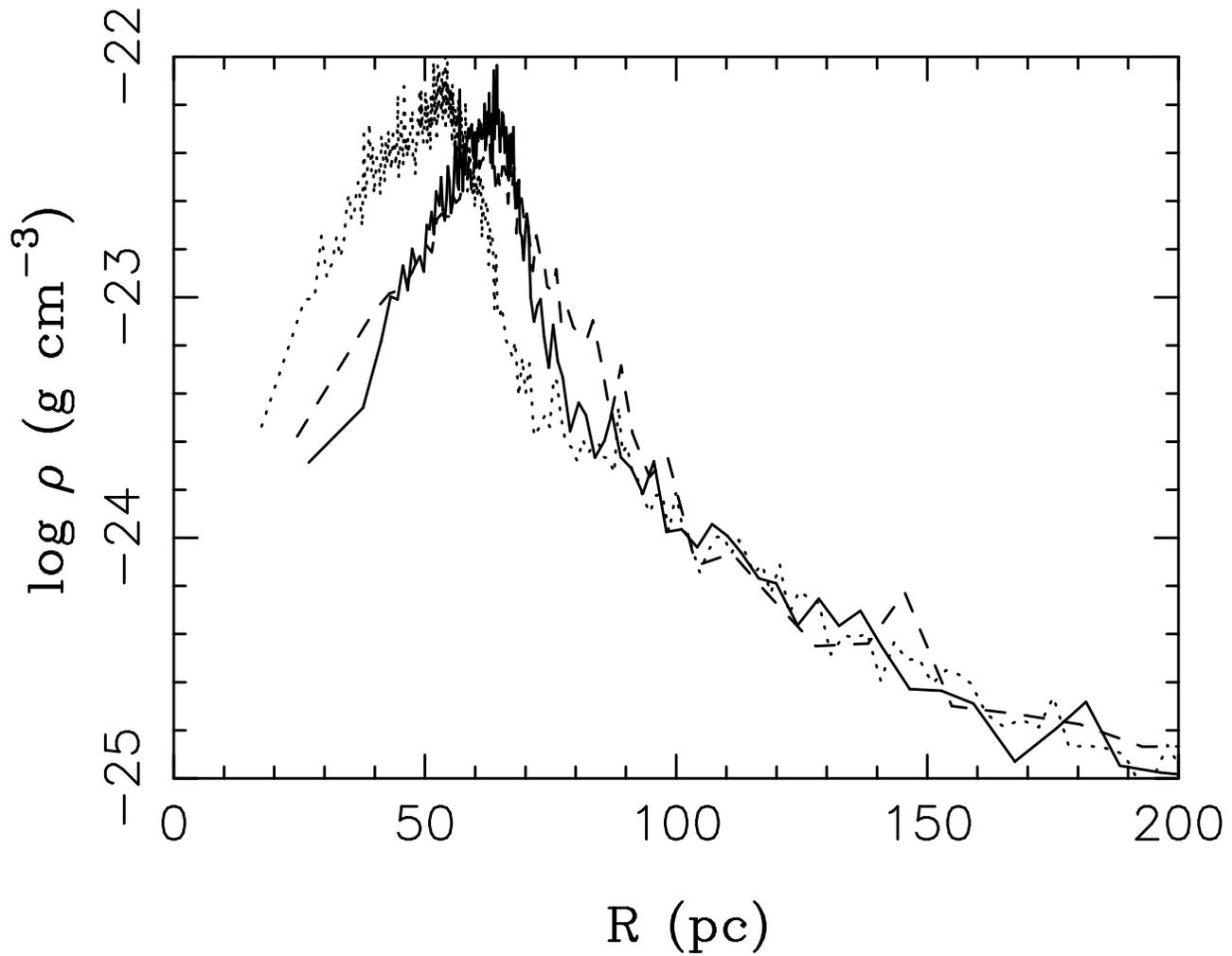}
\caption{
The comparison between the reference model (solid line) and other model.
The dashed line corresponds to the model with $N_i \sim$ 2500.
The dashed line corresponds to the model with $N_i \sim$ 10000.
\label{compare_1}
}
\end{figure}

First, we describe the dependence on the initial particle number. 
With larger (smaller) number of initial gas particles,
the initial masses of the gas particles are smaller (larger).
There is no strong dependence on the mass of the gas particles in
our star formation recipes because Eq. (\ref{star_mass}) dose not
include the mass of the gas particles.
However, the maximum density ($\rho_{\rm max}$) that can be represented
in the SPH method depends on the mass of the gas particles, 
which may produce weak dependence on $N_i$.
To confirm this, we evolve the model with $N_i = 10^4$.
The overall evolution is almost indistinguishable to the reference model.
The stellar mass at $t =$ 10 Myr becomes $\sim 1.2 \times 10^5$ \SM,
which is slightly smaller than $1.3 \times 10^5$ \SM~in the reference model.
For $N_i = 2500$ (a half of the reference model),
we obtain the stellar mass of $\sim 1.5 \times 10^5$ \SM.
In Figure \ref{compare_1}, we compare the gas density profiles at $t =$ 10 Myr
for different $N_i$.
The position of the shell for the model with $N_i = 10^4$ is different
from other models because of the smaller stellar mass at the same epoch.
However, all three models show the similar density profiles so that 
we conclude that the dependence on the initial gas particle number is week.
Thus, $N_i \sim$ 5000 is sufficient for the current numerical model.

Secondary, we describe the dependence on the size of
the feedback radius $R_f$. 
In our model, the cooling and the star formation are
suppressed for the gas particles in heating phase.
We therefore expect that the size of $R_f$ affects the star formation history.
With larger $R_f =$ 5 pc, the star formation rates are smaller and the
final stellar mass ($\sim 0.9 \times 10^5$ \SM)
is smaller than $1.3 \times 10^5$ \SM~ in the reference model ($R_f =$ 5 pc).
On the other hand, we obtain the stellar mass of
$\sim 1.9 \times 10^5$ \SM~ with smaller $R_f$ (= 2 pc).
Within the reasonable range of $R_f$ around 3pc,
the final stellar mass changes by a factor of 0.7 - 1.5.

\subsection{Summary of the numerical results}
The results of our calculations are summarized as follows
(see also Table \ref{table1}) :

\begin{enumerate}
\item
In all cases, the overall evolution is similar.
Initially, the star burst occurs in the central region. 
The central star burst is halted by the heating due to Type II supernovae.
The heating cause the expansion of the central region and 
forms a shell structure.
The subsequent star formation occurs in the shell.

\item
For the metal-free collapse case, the stellar mass at $t =$ 10 Myr correlates
with initial concentrations.
To obtain the star cluster as massive as globular clusters
($\sim 10^5$ \SM),
the initial concentration of the PGC must be large enough,
i.e., $R_i < 200$ pc.

\item
For the metal-rich collapse, the initial metallicity significantly
affects the evolution.
To obtain the star cluster as massive as globular clusters,
the initial metallicity must be as larger as [Fe/H] $\sim -2$.
If the initial metallicity is low ([Fe/H] $< -2$), very few stars form.

\item
This suggests that during the initial phase of the galaxy formation,
i.e., when the ISM contains little heavy element, 
the formation efficiency of the globular cluster is low.

\item
The results is not strongly affected by the initial number of gas particles.
$N_i \sim$ 5000 is sufficient for the current numerical model.
The dependence on the feedback radius is more evident.

\end{enumerate}

\section{Discussion}

\subsection{Star formation in the shell}
In all our numerical models, a shell-like gaseous structure is formed.
The formation of the shell-like structure of gases has also been reported in 
the simulations of the formation of dwarf elliptical galaxies
\citep{Mori_1997, Mori_1999}.
They argued that the difference in the density structure
between normal ellipticals (de Vaucouleurs law) and dwarf ellipticals
(exponential law) can be explained by the formation of
such a shell-like structure and the star formation in the shell.
The most crucial difference between the normal elliptical and the dwarf
is their mass, and the less massive galaxy is more strongly affected
by the energy inputs from stars.
Similar argument may be applicable to our numerical models.
In the case of the globular cluster formation, however, 
the mass is even smaller than the dwarf galaxies.
Owing to this fact, the effect of the energy inputs from stars
is more drastic so that the star formation in the shell is much less 
efficient than in the dwarf galaxies.
Also, the stars formed in the shell is not gravitationally bound
due to the outward velocity of the shell.
Thus, the bound globular cluster consists of the stars formed
before the shell formation.
The stellar density of such stars is not affected by the shell formation
and the star formation in the shell.

The chemical composition of stars in globular clusters is one of the
most crucial quantities to constrain the model for the globular cluster
formation.
The stars in a globular cluster have almost the same heavy elements
abundances \citep{Suntzeff_1993}.
This small dispersion in metallicity ($\sigma[{\rm Fe/H}]$) suggests that
the formation period of globular clusters is so short that
the stars in globular clusters can be regarded to form almost simultaneously
as shown in our numerical models.

\begin{figure}
\epsfbox[44 28 524 405]{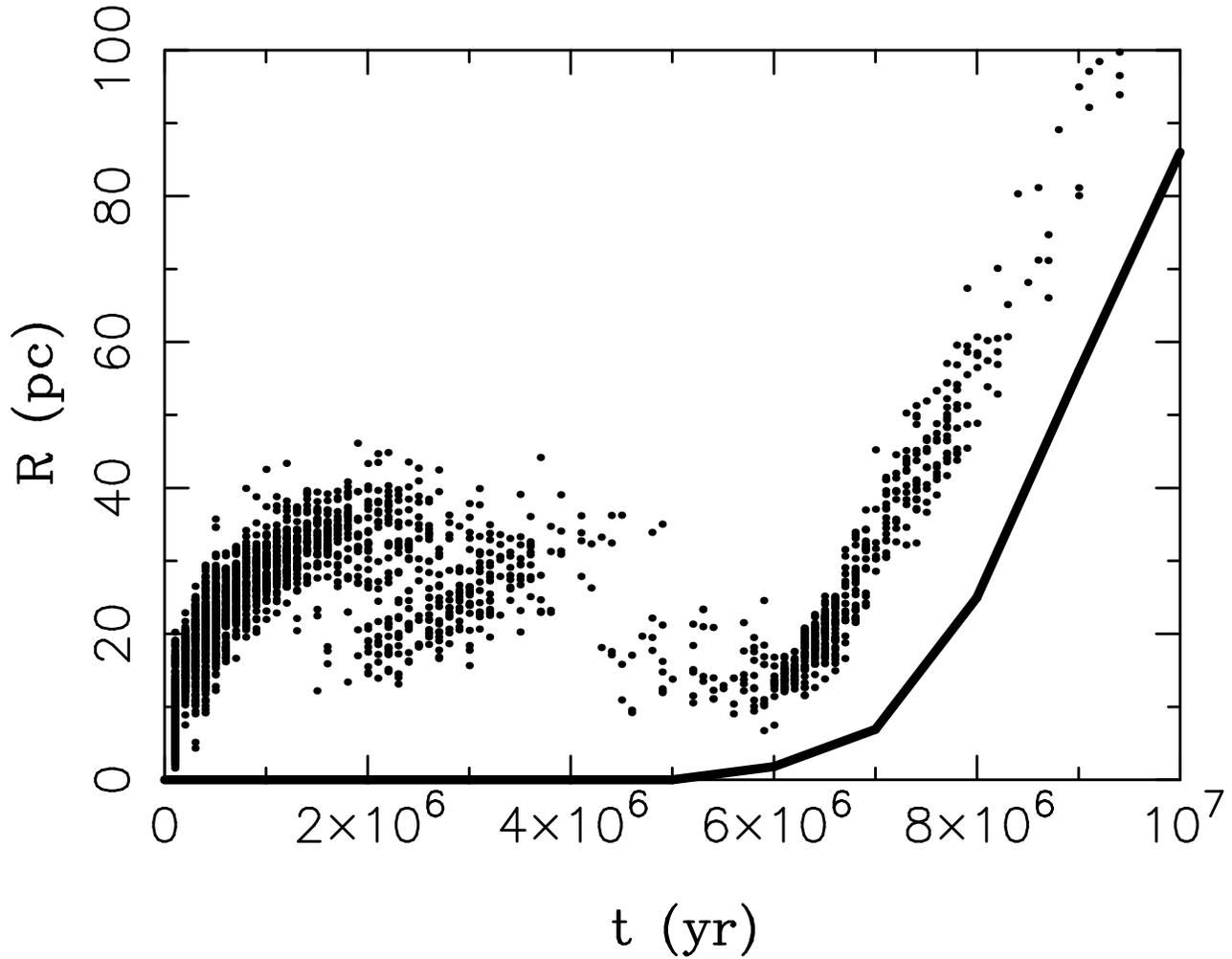}
\caption{
The radius (pc), at which stars form, as a function of time (yr)
for the metal-free collapse with $R_i =$ 150 pc.
\label{150_free_birth}
}
\end{figure}

\cite{Brown_1991, Brown_1995} suggested that
the second generation stars would form in the shell and 
the self-enrichment could occur there.
In our model, the star formation takes place in the shell
after $t =$ 5 Myr (see Figure \ref{150_free_birth}).
However, when the stars form in the shell, the shell has not been 
enriched with newly ejected heavy metal as shown in the solid line in Figure
\ref{150_free_birth}.
This line shows the radius of the metal-enriched region
defined as $r = (\Sigma r_i m_i)/(\Sigma m_i$)
by summing over the metal-enriched gas particles.
The metal-enriched region expands outwardly but does not reach 
the star forming region, which implies
that the self-enrichment dose not occur in our model
(we note that the star formation is suppressed in the gas particles
near the STARs in our numerical model as described in section 2.3.3).
In the present model, no external medium outside the cloud
is included, because of technical  difficulties in the SPH method 
(see, however, \citet{Nagasawa_Miyama_1987} for a possible improvement).
If there exists external medium outside the cloud, the density of the shell
would be higher and the star formation history would be different;
this possibility needs further study to confirm.

\subsection{Failed Globular clusters}
When a PGC is an initially metal-free cloud or
the initial metallicity of a PGC is low, 
the resulted mass and the central stellar density are
lower than the observed mass and density of the globular clusters.
Such ``failed'' globular clusters might be the field halo stars.
Another possibility is a open cluster.
Typical age of open clusters in our galaxy is $\tau < 10^9$ yr.
However, there also exist such old open clusters as $\tau > 10$ Gyr
\citep{Friel_1995}.
The age of the most old open cluster is comparable to the age of 
globular clusters. 
The central density of the ``failed'' globular cluster
is as low as the central density of typical open clusters.
Thus, the formation processes of such old open clusters may be the similar to
our model of globular cluster formation but starting from lower initial
concentration.
Very old open clusters might be the debris of ``failed'' 
globular clusters and there might have existed many more open clusters
at the formation of our galaxy.

\section{Conclusions}
\begin{figure}
\epsfxsize=0.7\hsize
\epsfbox[44 28 507 392]{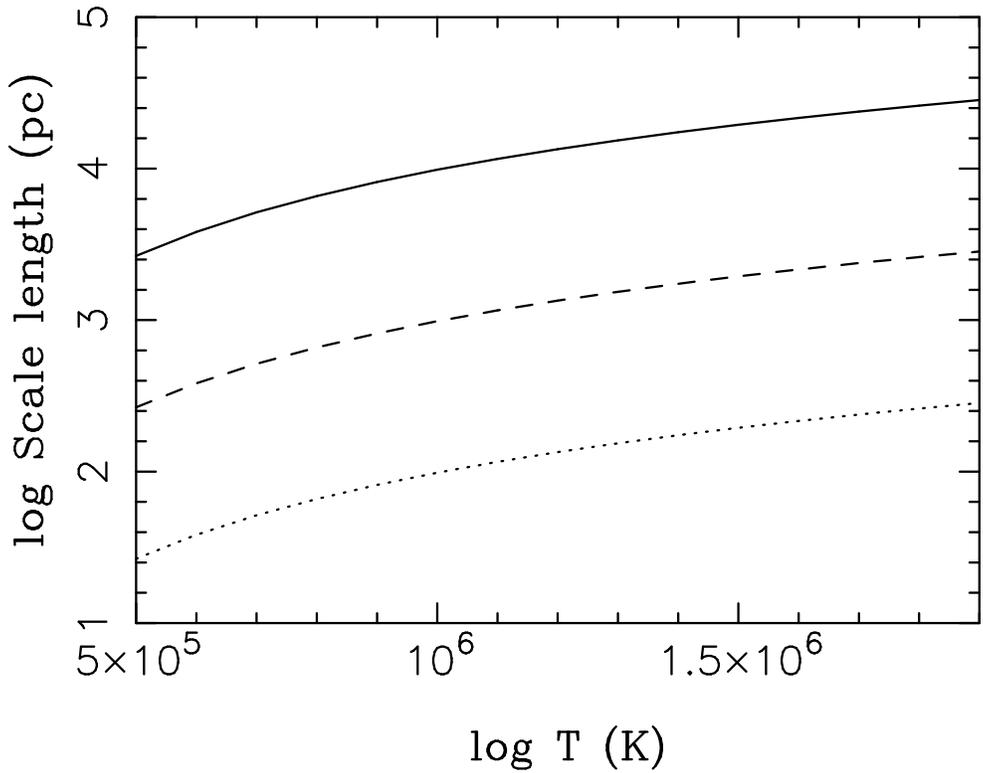}
\caption{
The estimated scale length of a proto-globular cloud
corresponding to the maximum growth rate in Eq. (\ref{chara}).
The solid, dashed, and dotted lines correspond to 
$\rho_0 = 1.0 \times 10^{-25}$, $1.0 \times 10^{-24}$,
and $1.0 \times 10^{-23}$ g cm$^{-3}$, respectively.
\label{m_n}
}
\end{figure}

For the processes of globular clusters formation, 
only the qualitative scenarios have been discussed previously 
\citep{Fall_Rees_1985, Lin_Murray_1996}. 
In this paper, we present the first attempt to simulate
the globular cluster formation with three-dimensional hydrodynamical method,
which includes the star formation and its feedback effects.
We assume that, in the collapsing galaxy, isothermal cold clouds
form through thermal condensations and become proto-globular clouds.
We obtain the size of proto-globular clouds by means of the linear
stability analysis (Figure \ref{m_n})
and compute the evolution of the inner region of the PGC starting from 
various initial radius $R_i$.
The results of our calculations are summarized as follows:

\begin{enumerate}
\item
In order for the globular cluster-like system to form from a metal-free PGC, 
the initial concentration of the PGC must be large enough.

\item
It is required that the metallicity of a PGC is high enough to produce
the globular cluster-like system.
In our calculations, the required metallicity is estimated to be
[Fe/H] $\geq -2$.

\item
In all cases, the shell like structure of the gas forms.
Although the star formation occurs in the shell,
the self-enrichment is not seen to occur.
\end{enumerate}

Based on the earlier qualitative works and the present quantitative results, 
the processes of globular clusters formation in the proto-galaxy
can be understood as follows:

\begin{enumerate}
\item
In the collapsing proto-galaxy, the first generation stars of
$M \sim 10^2$ \SM~ form due to the efficient cooling by hydrogen molecules.
Such population III stars eject the gas with heavy elements and
chemically contaminate the proto-galaxy gases.

\item
Such first generation stars radiate dissociative photons, and
the entire proto-galaxy is settled down to ionization equilibrium.

\item
With equilibrium cooling, the density perturbation grows due to
thermal instability. and forms an isothermal cloud
with a density structure of $\propto r^{-2}$.
Such clouds are the proto-cloud of globular clusters.

\item
When the density of the PGC becomes high enough, the burst of star formation 
occurs.
Some high density clouds produce the globular clusters, and others may
produce the field stars and/or the halo open clusters.

\item
During the formation of the galaxy, 
the formation efficiency of the globular cluster become significantly large
when the metallicity of the PGC become as large as [Fe/H] $\sim -2$.

\end{enumerate}

As mentioned in section 5.1, we do not take account of
the effect of the external medium in the present calculations.
If we include the effect of the external medium, 
the structure of the shell may be different and further star formation may
occur in that shell. \citet{Brown_1995} suggested that only a few supernovae
per Myr is sufficient to reverse the contraction of the 10$^6$ \SM~ cloud.
Such an effect of the external medium needs to be studied.

The globular cluster formation is considered to occur not only 
in the proto-galaxies but also in the present galaxies, e.g., LMC, SMC,
and interacting galaxy NGC4038/NGC4039.
In such environment, different formation processes would take place, 
which correspond to the tertiary models \citep{Kumai_1993}.
In the case of NGC4038/NGC4039 (the Antennae galaxies),
many globular clusters are being produced by galaxy merging.
Detail formation processes of globular clusters in such systems
needs to be studied.
Also, the effect of the formation of many globular clusters
on the galaxy merging process needs a further study.

\bigskip

We would like to thank the anonymous referee
for the valuable suggestions to improve the manuscript.
Also, we would like to thank G. Mathews, J. Truran, T. Shigeyama
and S. Zwart for useful discussion and comments.
This work has been supported by JSPS Research Fellowships for 
Young Scientists (7664),
and in part by the grant-in-Aid for COE Scientific Research (07CE2002)
of Ministry of Education, Science, Culture, and Sports of Japan.

\clearpage

\begin{deluxetable}{ccccc}
\tablecaption{The summary of the numerical results
\label{table1}
}
\tablehead{
\colhead{$R_i$$^{\rm a}$} &
\colhead{[Fe/H]$^{\rm b}$} &
\colhead{Mass$^{\rm c}$} &
\colhead{$\rho_{\rm c}$$^{\rm d}$} &
\colhead{$\sigma_{\rm c}$$^{\rm e}$}
}
\startdata
150 pc & no metal & $1.0 \times 10^5$ \SM & 10$^{2.5}$ \SM~ pc$^{-3}$ & 2.98 km s$^{-1}$ \\
200 pc & no metal & $5.1 \times 10^4$ \SM & 10$^{1.1}$ \SM~ pc$^{-3}$ & 1.87 km s$^{-1}$ \\
300 pc & no metal & $5.5 \times 10^4$ \SM & 10$^{1.4}$ \SM~ pc$^{-3}$ & 3.66 km s$^{-1}$ \\
300 pc & $ 0$     & $1.0 \times 10^5$ \SM & 10$^{2.0}$ \SM~ pc$^{-3}$ & 3.21 km s$^{-1}$ \\
300 pc & $-1$     & $1.5 \times 10^5$ \SM & 10$^{2.0}$ \SM~ pc$^{-3}$ & 2.33 km s$^{-1}$ \\
300 pc & $-2$     & $1.0 \times 10^5$ \SM & 10$^{1.7}$ \SM~ pc$^{-3}$ & 2.68 km s$^{-1}$ \\
300 pc$^{f}$ & $-3$     & $2.9 \times 10^3$ \SM &  &  \\
\enddata
\tablecomments{
$^{\rm a}$The initial radius of the PGC. \\     
$^{\rm b}$The initial metallicity ([Fe/H]) of the PGC.\\     
$^{\rm c}$The bound stellar mass at $t =$ 10 Myr.\\     
$^{\rm d}$The central stellar density at $t =$ 10 Myr.\\     
$^{\rm e}$The central stellar velocity dispersion at $t =$ 10 Myr.\\     
$^{\rm f}$ $\rho_{\rm c}$ and $\sigma_{\rm c}$ is omitted of this case.
}
\end{deluxetable}

\begin{deluxetable}{ccc}
\tablecaption{The dependence on numerical parameters
\label{table2}
}
\tablehead{
\colhead{$N_i$$^{\rm a}$} &
\colhead{$R_f$$^{\rm b}$} &
\colhead{Mass$^{\rm c}$}
}
\startdata
5000  & 3 pc & $1.5 \times 10^5$ \SM \\
2500  & 3 pc & $1.5 \times 10^5$ \SM \\
10$^4$& 3 pc & $1.2 \times 10^5$ \SM \\
5000  & 2 pc & $1.9 \times 10^5$ \SM \\
5000  & 5 pc & $8.6 \times 10^4$ \SM \\
\enddata
\tablecomments{
$^{\rm a}$Used initial gas particle number.\\     
$^{\rm b}$The feedback radius.\\     
$^{\rm c}$The stellar mass at $t =$ 10 Myr.\\     
}
\end{deluxetable}

\clearpage

\clearpage

\clearpage

\clearpage

\clearpage

\clearpage

\clearpage

\begin{thebibliography}{}

\bibitem[Abbot (1982)]{Abbot_1982}
Abbot, D.C., 1982, ApJ, 263, 723

\bibitem[Brown, Burkert \& Truran (1991)]{Brown_1991}
Brown, J.H., Burkert, \& A., Truran, J.W., 1991, ApJ, 376, 115

\bibitem[Brown, Burkert \& Truran (1995)]{Brown_1995}
Brown, J.H., Burkert, A., \& Truran, J.W., 1995, ApJ, 440, 666

\bibitem[David, Forman, \& Jones (1990)]{David_1990}
David, L.P., Forman, W., \& Jones, C., 1990, ApJ, 459, 29

\bibitem[Dubath, Meylan \& Mayor (1997)]{Dubath_1997}
Dubath, P., Meylan, G., \& Mayor, M., A\&A, 324, 505

\bibitem[Ebert (1955)]{Ebert_1955}
Ebert, R., 1955, Zs. Ap., 37, 217

\bibitem[Elmergreen et al. (1999)]{Elmergreen_1999}
Elmergreen, B.G., Efremov, Y., Pudritz, R.E., \& Zinnecker, H., 1999,
astro-ph/9903136

\bibitem[Fall \& Rees (1985)]{Fall_Rees_1985}
Fall, S.M., \& Rees, M.J., 1985, ApJ, 298, 18

\bibitem[Fall \& Rees (1987)]{Fall_Rees_1987}
Fall, S.M., \& Rees, M.J., 1987, in IAU Symposium 126,
Globular cluster systems in galaxies, eds. Grindlay, J. \& Philip, D.
(Dordrecht: Kluwer), 323

\bibitem[Field (1965)]{Field_1965}
Field, G.B., 1965, ApJ, 142, 531

\bibitem[Friel (1995)]{Friel_1995}
Friel, E.D., 1995, Annu. Rev. Aston. Astrophys., 33, 381

\bibitem[Friedli \& Benz (1995)]{Friedli_Benz_1995}
Friedli, D., \& Benz, W., 1995, A\&A, 301, 649

\bibitem[Gingold \& Monaghan (1977)]{Gingold_Monaghan_1977}
Gingold, A., \& Monaghan, J., 1977, MNRAS, 181, 37

\bibitem[Harris (1991)]{Harris_1991}
Harris, W.E, 1991, Annu. Rev. Aston. Astrophys., 29, 543

\bibitem[Ho \& Filippenko (1996)]{Ho_Filippenko_1996}
Ho, L.C., \& Filippenko, A.V., 1996, ApJL, 466, 83

\bibitem[Katz (1992)]{Katz_1992}
Katz, N., 1992, ApJ, 391, 502

\bibitem[Katz, Weinberg, \& Hernquist (1995)]{Katz_1995}
Katz, N., Weinberg, D.H., \& Hernquist, L., 1996, ApJS, 105, 19 

\bibitem[Kumai, Basu, \& Fujimoto (1993)]{Kumai_1993}
Kumai, Y., Basu, B., \& Fujimoto, M., 1993, ApJ, 404, 144

\bibitem[Leitherer, Robert, \& Drissen (1992)]{Leitherer_1992}
Leitherer, C., Robert, C., \& Drissen, L., 1992, ApJ 401, 596

\bibitem[Lin \& Murray (1996)]{Lin_Murray_1996}
Lin, D.N.C., \& Murray, S.D., 1996, in IAU Symposium 174,
Dynamical evolution of star clusters, eds. Hut., P. \& Makino, J.
(Dordrecht: Kluwer), 283

\bibitem[Lucy (1977)]{Lucy_1977}
Lucy, L., 1977, AJ, 82, 1013

\bibitem[McCrea (1957)]{McCrea_1957}
McCrea, W.H., 1957, MNRAS, 117, 562

\bibitem[Meerson (1996)]{Meerson_1996}
Meerson, B., 1996, Reviews of Modern Physics, 68, 1, 215

\bibitem[Meylan, \& Heggie (1997)]{Meylan_Heggie_1997}
Meylan, G., Heggie, D.C., 1997, A\&AR, 8, 1

\bibitem[Mori et al. (2000)]{Mori_2000}
Mori, M., Nakasato, N., Yoshii, Y., \& Nomoto, K., 2000, in preparation

\bibitem[Mori et al. (1997)]{Mori_1997}
Mori, M., Yoshii, Y., Tsujimoto, T., \& Nomoto, K., 1997, ApJL, 478, 21

\bibitem[Mori, Yoshii, \& Nomoto (1999)]{Mori_1999}
Mori, M., Yoshii, Y., \& Nomoto, K., 1999, ApJ, 511, 585

\bibitem[Murray \& Lin (1993)]{Murray_Lin_1993}
Murray, S.D., \& Lin, D.N.C., 1993, in ASP Conf. Ser. Vol. 48,
The Globular Cluster-Galaxy Connection, eds. Smith, G.H., \& Brodie, J.P.
(San Francisco: Astronomical Society of the Pacific), 738

\bibitem[Nagasawa \& Miyama (1987)]{Nagasawa_Miyama_1987}
Nagasawa, M., \& Miyama, S.M., 1987, Prog. Theor. Phys., 78, 1250

\bibitem[Nakasato et al. (1996)]{Nakasato_1996}
Nakasato, N., Mori, Tsujimoto, T., Mathews, G., \& Nomoto, K., 1996,
in IAU Symposium 174, Dynamical Evolution of Star Clusters, 
eds. Hut, P., \& Makino, J.(Dordrecht: Kluwer), 397

\bibitem[Nakasato, Mori, \& Nomoto (1997)]{Nakasato_1997}
Nakasato, N., Mori, M., \& Nomoto, K., 1997, ApJ, 484, 608

\bibitem[Nakasato, Mori, \& Nomoto (1999)]{Nakasato_1999}
Nakasato, N., Mori,  \& Nomoto, K., 1999,
in IAU Symposium 187, Cosmic Chemical Evolution, 
eds. Truran, J.W., \& Nomoto, K.(Dordrecht: Kluwer), in press

\bibitem[Navarro \& White (1993)]{Navarro_White_1993}
Navarro, J.F., \& White, S.D.M., 1993, MNRAS, 265, 271

\bibitem[Nomoto et al. (1997)]{Nomoto_1997}
Nomoto, K., Hashimoto, M., Tsujimoto, T., Thielemann, F.-K.,
Kishimoto, N., Kubo, Y., \& Nakasato, N., 1997, Nucl. Phys. A., 616, 79c

\bibitem[Osterbrock (1974)]{Osterbrock_1974}
Osterbrock, D.E., 1974, Astrophysics of Gaseous Nebulae
(San Fransisco: W. H. Freeman and Company)

\bibitem[Palmer \& Papaloizou (1987)]{Palmer_Papaloizou_1987}
Palmer, P.L., \& Papaloizou, J., 1987, MNRAS, 224, 1043

\bibitem[Parker (1953)]{Parker_1953}
Parker, E.N., 1953, ApJ, 117, 431

\bibitem[Peebles \& Dicke (1968)]{Peebles_Dicke_1968}
Peebles, P.J.E., \& Dicke, R.H., 1968, ApJ, 154, 891

\bibitem[Schmidt (1959)]{Schmidt_1959}
Schmidt, M., 1959, ApJ, 129, 243

\bibitem[Shapiro \& Kang (1987)]{Shapiro_Kang_1987}
Shapiro, P.R., \& Kang, H., 1987, ApJ, 318, 32

\bibitem[Spitzer(1978)]{Spitzer_1978}
Spitzer, L., Jr., 1978, Physical Processes in the Interstellar Medium
(New York: Wiley Interscience)

\bibitem[Steinmetz \& M\"uller (1994)]{Steinmetz_Muller_1994}
Steinmetz, M., \& M\"uller, E., 1994, A\&A, 281, 97

\bibitem[Sugimoto et al. (1990)]{Sugimoto_1990}
Sugimoto, D., Chikada, Y., Makino, J., et al., 1990, Nature, 345, 33

\bibitem[Suntzeff (1993)]{Suntzeff_1993}
Suntzeff, N., 1993, in ASP Conf. Ser. Vol. 48,
The Globular Cluster-Galaxy Connection, eds. Smith, G.H., \& Brodie, J.P.,
(San Francisco: Astronomical Society of the Pacific), 167 

\bibitem[Sutherland \& Dopita (1993)]{Sutherland_Dopita_1993}
Sutherland, R.S., \& Dopita, M.A., 1993, ApJS, 88, 253

\bibitem[Thacker et al. (1998)]{Thacker_1998}
Thacker, R.J., Tittley, E.R., Pearce, F.R., Couchman, H.M.P., \& Thomas, P.A.,
1998, astro-ph/989221

\bibitem[Tsujimoto et al. (1996)]{Tsujimoto_1996}
Tsujimoto, T., Nomoto, K., Yoshii, Y., Hashimoto, M., Yanagida, S., \& 
Thielemann, F.-K., 1996, MNRAS, 277, 945

\end{thebibliography}
\end{document}